# Near surface defects: Cause of deficit between internal and external open-circuit voltage in solar cells

Mohit Sood*[1], Aleksander Urbaniak[2], Christian Kameni Boumenou[1], Thomas Weiss[1], Hossam Elanzeery[1], Finn Babbe[1,3], Florian Werner[1,4], Michele Melchiorre[1], Susanne Siebentritt[1]

(E-mail address: mohit.sood@uni.lu)

[1]*Department of Physics and Materials Science, University of Luxembourg, Belvaux, L-4422, Luxembourg*

[2]*Faculty of Physics, Warsaw University of Technology, Koszykowa 79, Warszawa 00-662, Poland*

[3]*Chemical Sciences Division, Joint Center for Artificial Photosynthesis, Lawrence Berkeley National Laboratory, Berkeley, USA*

[4]*Hydrosat, 9, rue du Laboratoire, L-1911 Luxembourg*

**Abstract**

The presence of interface recombination in a complex multilayered thin-film solar structure causes a disparity between the internal open-circuit voltage ($V_{OC,in}$), measured by photoluminescence, and the external open-circuit voltage ($V_{OC,ex}$) *i.e.* an additional $V_{OC}$ deficit. Higher $V_{OC,ex}$ value aim require a comprehensive understanding of connection between $V_{OC}$ deficit and interface recombination. Here, a deep near-surface defect model at the absorber/buffer interface is developed for copper indium di-selenide solar cells grown under Cu-excess conditions to explain the disparity between $V_{OC,in}$ and $V_{OC,ex}$.. The model is based on experimental analysis of admittance spectroscopy and deep-level transient spectroscopy, which show the signature of deep acceptor defect. Further, temperature-dependent current-voltage measurements confirm the presence of near surface defects as the cause of interface recombination. The numerical simulations show strong decrease in the local $V_{OC,in}$ near the absorber/buffer interface leading to a $V_{OC}$ deficit in the device. This loss mechanism leads to interface recombination without a reduced interface bandgap or Fermi level pinning. Further, these findings demonstrate that the $V_{OC,in}$ measurements alone can be inconclusive and might conceal the information on interface recombination pathways, establishing the need for complementary techniques like temperature dependent current-voltage measurements to identify the cause of interface recombination in the devices.

*corresponding author: mohit.sood@uni.lu

## Introduction

Open-circuit voltage ($V_{OC}$), a key factor for the efficiency of a solar cell, is measured by either electrical or optical techniques. Electrical measurements, particularly current-voltage measurements give the measure of external open-circuit voltage ($V_{OC,ex}$) of a device, whereas, optical measurements particularly calibrated photoluminescence (PL) provide the measure of the internal open-circuit voltage ($V_{OC,in}$) or quasi-Fermi level splitting (qFLs). The $V_{OC,in}$ (qFLs) is calculated from the ratio of total radiative recombination flux of the device to the flux of injected photons. It is generally measured *via* one sun calibrated PL measurement (in order to compare it to AM 1.5 G illuminated solar cell $V_{OC,ex}$), and translates to the energetic difference between the hole quasi-Fermi level ($F_h$) and electron quasi-Fermi level ($F_e$) in the bulk.[1] Moreover, $V_{OC,in}$ provides a direct measure of the bulk quality of an absorber. While, the $V_{OC,ex}$ measured in a current-voltage (I-V) measurement under one sun illumination is the energetic difference between the $F_h$ at the hole contact and the $F_e$ at the electron contact. The $V_{OC,ex}$ takes into account the interfaces and contacts as well, and is a device related parameter. Hence, $V_{OC,ex}$ is a metric that represent the overall quality of the device. In order to translate optical quality of the absorber into electrical efficiency *i.e.* $V_{OC,ex}$ it is essential to have a constant qFLs throughout the device structure.[2-4]

Thin films solar cells comprise of a complex multilayer structure consisting of absorber, charge transport layer etc., each of which individually affect the qFLs and could be a source of a gradient in qFLs. This often leads to a deficit between internal and external $V_{OC}$ *i.e.* $V_{OC,in} – V_{OC,ex}$. The deficit can be observed in thin film solar cells such as $Cu(In,Ga)(Se,S)_2$,[5,6] CdTe,[7] perovskite,[4,8,9] and is associated to interface recombination in the device.[4,9-13] Identifying the source of interface recombination and the underlying qFLs gradient is crucial for achieving higher efficiency in these devices and enabling better understanding of device physics. The mismatch of the energy bands at interface between absorber and charge transport layer,[4,14,15] and Fermi-level pinning are the two commonly evoked models to explain why and more so, in which case interface recombination dominates.[16-19]

Researchers employ qFLs measurements for quantifying interface recombination and determining the quality of surface passivation after charge transport layer deposition or post deposition treatment.[4,20] Though qFLs measurements provide significant information regarding non-radiative recombination in the bulk, it fails to capture the details of interface processes especially in devices dominated by IF recombination.[21] The photoluminescence intensity increases exponentially with the qFLs. Thus in the case of a qFLs gradient, photoluminescence will always detect the highest qFLs and will not indicate the gradient.[15] Therefore, temperature dependent $V_{OC,ex}$ measurements are required to unravel the presence of interface recombination in the device and thus provide necessary information to understand the full extent of the non-radiative interface recombination losses in the device.[22]

Here, with the help of copper indium diselenide (CISe), a chalcogenide photovoltaic absorber material, we develop a comprehensive model for understanding the interface $V_{OC}$ deficit by probing the effect of near-surface defects on $V_{OC,in}$ and $V_{OC,ex}$ of the CISe device. We choose CISe for studying the interface $V_{OC}$ deficit, since CISe absorbers grown under Cu-excess conditions (addressed as Cu-rich throughout this work with as grown [Cu]/[In] > 1) and under In-excess (addressed as Cu-poor with as grown [Cu]/[In] < 1) growth conditions result in similar $V_{OC,in}$ with completely different $V_{OC,ex}$, and therefore different interface $V_{OC}$ deficit.[10,23,24] Moreover, instead of the commonly used $Cu(In,Ga)Se_2$ compounds that have bandgap graded absorber layers,[25] the ternary CISe compound allows to reduce the amount of free variables and redundant complexity in our model. This makes CISe an ideal case study to investigate the cause of the interface $V_{OC}$ deficit in thin film solar cells.

We vary the interface defect density by treating Cu-rich absorbers with different solutions namely, aqueous KCN, aqueous bromine ($Br_{aq.}$), aqueous zinc ($Zn_{aq.}$), sulfur (S) and cadmium (Cd) solution, as well as by depositing a Zn(O,S) buffer. With the help of admittance spectroscopy (AS), deep-level transient spectroscopy (DLTS) and temperature-dependent current-voltage (I-V-T) measurements we probe the impact of these treatments. The study identifies the role of defects near (not at) the interface, which was hitherto not discussed. Furthermore, we scrutinize the limitations of $V_{OC,in}$ (qFLs) measurements alone in characterizing interface recombination and the necessity of temperature-dependent $V_{OC,ex}$ measurements. Using numerical modelling, we establish a model based on strong sub-surface defects, which demonstrates an interface $V_{OC}$ deficit for an interface with favorable band alignment and no Fermi level pinning. The model is experimentally endorsed and provides insights on the origin and nature of these sub-surface defects in CISe solar cells.

### Experimental observations Cu-rich vs Cu-poor CISe solar cell

Before building a comprehensive model, it is necessary to look at the optical and electrical characteristics of CISe solar cells prepared using absorbers grown under Cu-rich and Cu-poor growth conditions. Throughout this work $V_{OC,in}$ will be used to define the qFLs, and the deficit between $V_{OC,in}$ and $V_{OC,ex}$ will be referred to as interface $V_{OC}$ deficit, unless stated otherwise. Figure 1a shows typical I-V characteristics of Cu-rich and Cu-poor devices. Both devices are processed in a similar manner *i.e.* with same buffer (CdS) and window layer (i-ZnO+AZO), deposited with identical process parameters. The Cu-rich device exhibits a lower $V_{OC,ex}$ compared to Cu-poor device, even though both absorbers have almost the same $V_{OC,in}$ values [table in Figure 1b]. The $V_{OC,in}$ is measured with the help of calibrated PL measurements which were performed using our own lab-built system with continuous wave 663 nm diode laser as an excitation source. For extracting $V_{OC,in}$, samples covered with buffer layer on top are illuminated with laser and PL is measured. Intensity and spectral corrections is then applied to the raw data to determine $V_{OC,in}$, the entire procedure details can be found in reports[23,26]. An exemplary PL spectra is presented in Figure S1. As a consequence, Cu-rich devices suffer from a high interface $V_{OC}$ deficit (~130 mV), similar to previous data on $Cu(In,Ga)Se_2$.[23] This is significantly higher than the one in Cu-poor device (~20 mV) or in fully optimized devices (~10 mV).[27] This IF $V_{OC}$ deficit is clearly associated to IF recombination being the dominant recombination path in the device as revealed from $V_{OC,ex}$ measurements at different temperatures [Figure 1c]. The activation energy ($E_a$) of the saturation current density is obtained from extrapolation of $V_{OC,ex}$ to 0 K.[19] For Cu-rich devices, $E_a$ is always lower than the bulk bandgap ($E_G$) and is associated to the presence of deep defects.[28,29] Whereas, in Cu-poor devices $E_a$ extrapolates to the $E_G$ and hence, IF recombination does not limit $V_{OC,ex}$. Furthermore, an 'S shape' in the first quadrant is observed at lower temperatures in Cu-rich devices, which is not present in Cu-poor device [Figure 1d]. This roll-over in the first quadrant indicates a barrier for the forward current.[30] Problematic interface properties often lead to an S-shape in the fourth quadrant, which indicates an extraction barrier for the photocurrent.[31] Thus, a model will be valid only if it can successfully reproduce the three observations made for Cu-rich devices: (i) a large interface $V_{OC}$ deficit, (ii) an $E_a$ of the saturation current smaller than the $E_G$ and (iii) a 'S shape' in the first quadrant. However, to build a reliable model, we will first probe the characteristics of the deep defect that has been speculated to be the cause of all these issues in Cu-rich $CuInSe_2$.[32]

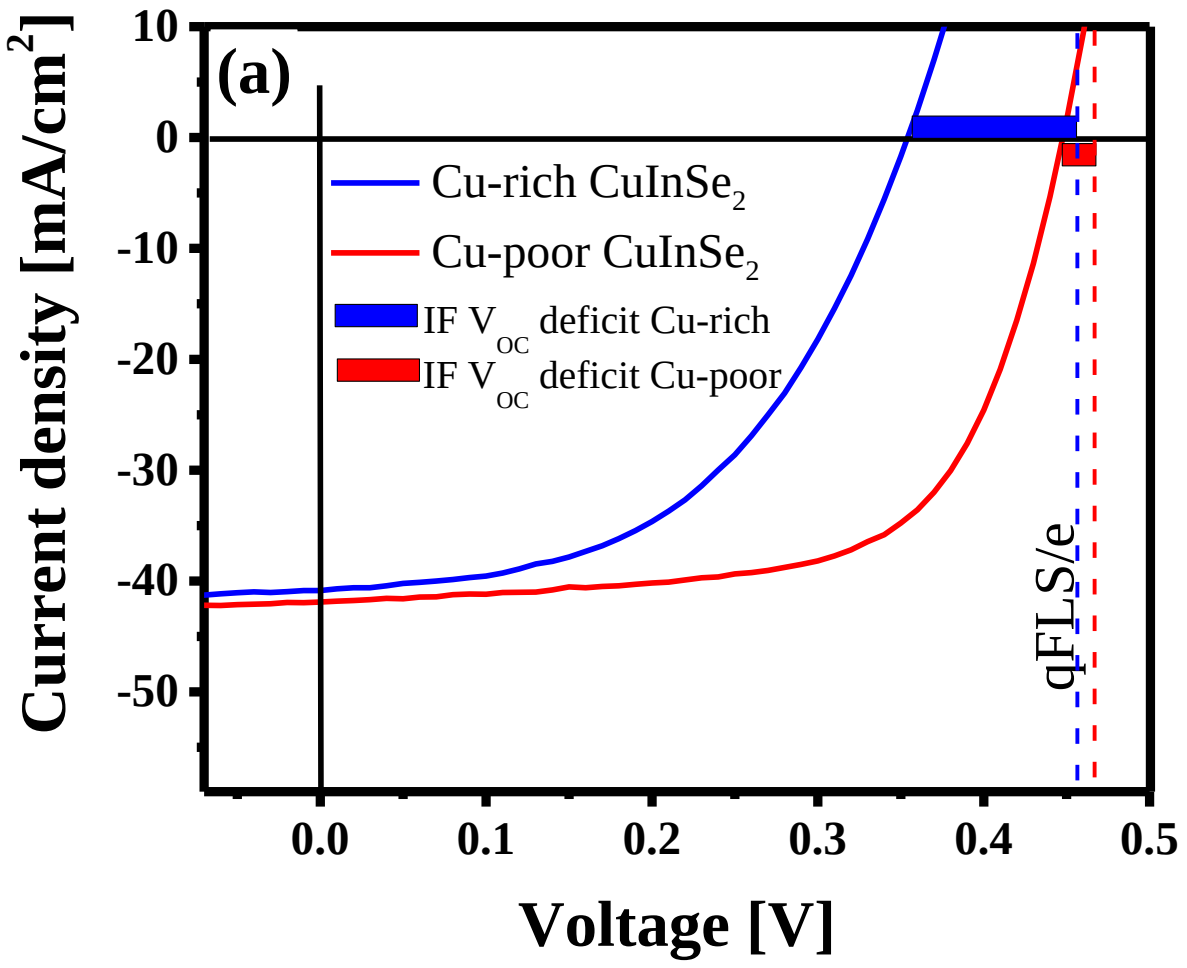

(b)

| | η [%] | $J_{sc}$ [mA/cm²] | FF [%] | $V_{OC,ex}$ [mV] | $V_{OC,in}$ [mV] |
|---|---|---|---|---|---|
| Cu rich | 7.2 | 42 | 50.5 | 340 | 457 |
| Cu poor | 12.7 | 44.5 | 62.4 | 456 | 466 |

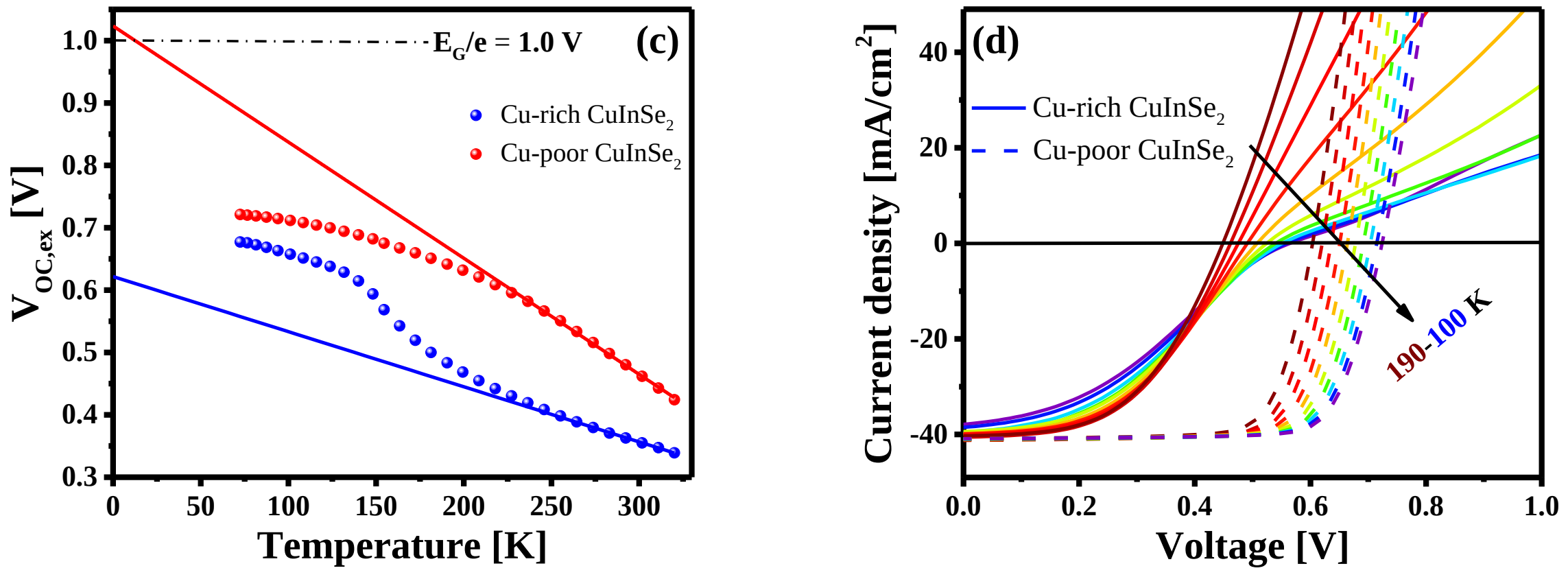

***Figure 1.*** Comparison of Cu-rich vs Cu-poor CISe device (a) and (b) I-V curve and characteristics, the comparison of $V_{OC,ex}$ and $V_{OC,in}$ shows a high interface $V_{OC}$ deficit for Cu-rich devices. The blue and the red bar show the interface $V_{OC}$ deficit for Cu-rich and Cu-poor device. (c) $V_{OC,ex}$ as a function of temperature, extrapolation to 0 K activation energy of saturation current density (d) I-V as a function of temperature, a 'S shape' at lower temperatures is observed in Cu-rich devices.

## Origin and characteristics of the deep defects

Despite its superior morphological and optoelectronic properties, the device performance of CISe absorbers grown under Cu-rich conditions is inferior to its Cu-poor counterpart.[11] This is due to the necessary KCN etching step required to remove the secondary $Cu_{2-x}Se$ phase. The etching results in high concentration >$10^{16}$ $cm^{-3}$ of deep defects (~200 meV) in Cu-rich CISe absorbers.[32,33] However, it is unknown whether the defect originates specifically from the KCN etching or from the etching process of secondary phase independent of the etchant used. To investigate this, Cu-rich CISe solar cells are prepared using two different etching solutions: 10% aqueous KCN solution (for reference) and 0.16 % mM aqueous Br solution. The impact of etching on the defect structure is investigated by measuring admittance spectroscopy (AS). Figure 2a shows exemplary AS measurements for KCN etched Cu-rich CISe solar cell. The spectra exhibit a capacitance step in the temperature range 190-100 K. The corresponding frequency derivatives of the AS spectra demonstrate broad asymmetric peaks (Figure 2b). These broad peaks are a peculiar feature always present in the AS spectra corresponding to the ~200 meV defect.[32] In comparison, the AS of aqueous Br etched Cu-rich CISe solar cell also exhibits a similar capacitance step (Figure S2b). More importantly, the inflection frequencies of AS of this device plotted together with that of the KCN etched device in an Arrhenius plot lie very close to each other, with activation energies around 200 meV. This indicates presence of a similar capacitance response in both the devices. In supplement to these results, a device prepared from a Br etched absorber also has the same $E_a$ of the saturation current as the KCN etched device, significantly lower than $E_G$ (Figure S3), signifying the presence of prevailing interface recombination. Thus, both results, the presence of similar capacitance step with a similar activation energy and the presence of interface recombinations, confirm the existence of the deep defect independent of the etchant used to remove the $Cu_{2-x}Se$ phase. This suggests that the ~200 meV defect is an intrinsic defect originating from the removal of the secondary phase from Cu-rich CISe films, as suggested in the literature.[34]

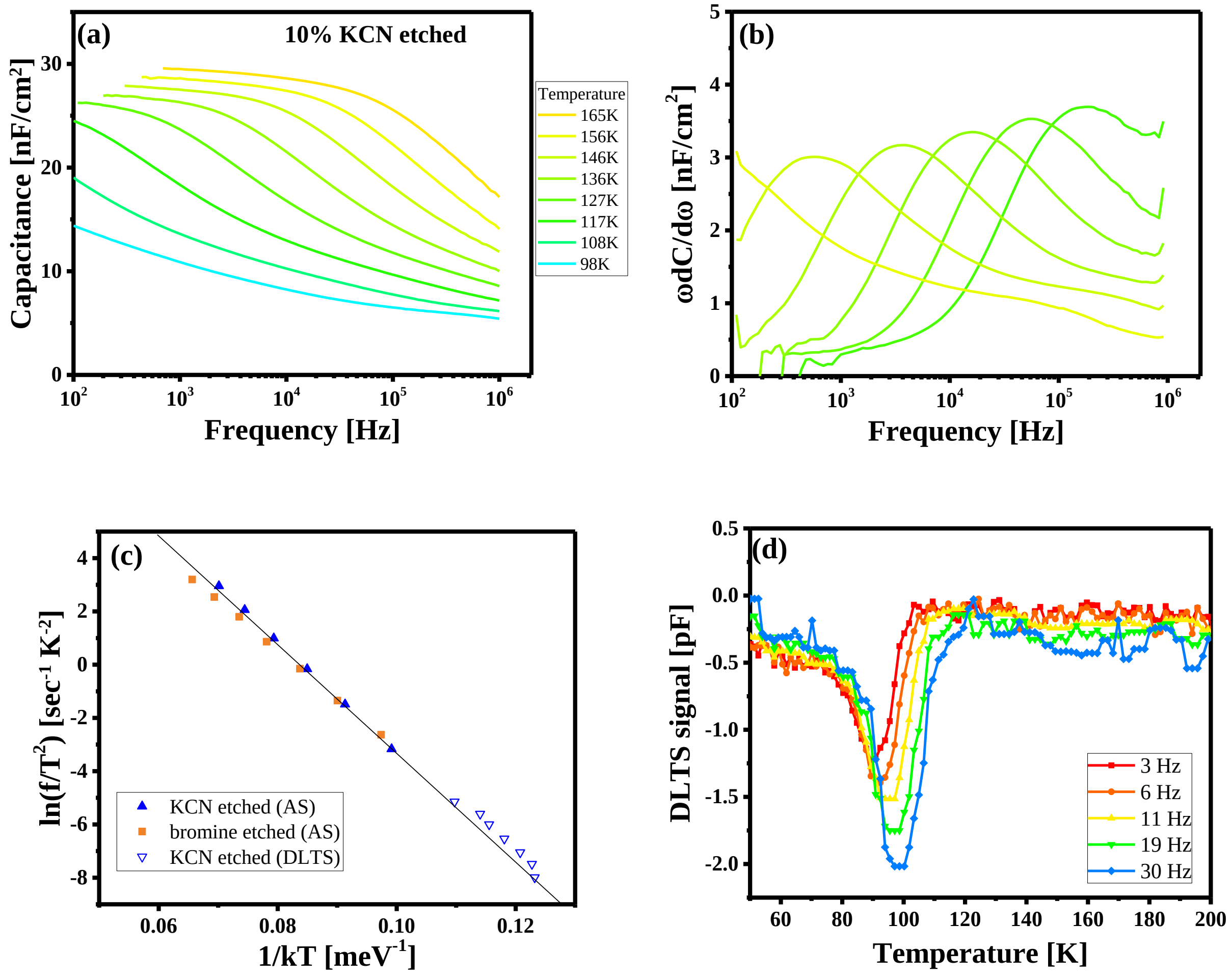

***Figure 2.*** (a) Admittance spectra of Cu-rich CISe solar cell prepared from absorbers etched with 10% KCN solution in a Schottky junction device. (b) ωdC/dω plot of corresponding admittance spectra, the peaks are broad and asymmetric. (c) Arrhenius plot of measured admittance (closed symbols) and DLTS (open symbols) measurements of CISe Schottky junction devices prepared with KCN etched and with bromine etched absorbers. (d) The DLTS signals of the KCN etched CISe Schottky junction device.

Although AS provides the defect activation energy, it does not yield the defect nature. Therefore, to investigate whether the defect is acceptor or donor in nature, DLTS is measured on KCN etched

CISe Schottky-devices (Figure 2c). For the measurement, the device was kept at -1V bias followed by a +1V voltage pulse and the capacitance transient was measured. Figure 2d shows the DLTS results for a chosen rate window alongside with the corresponding Arrhenius plot in Figure 2c. The peak in the DLTS spectrum is negative, which is a fingerprint of emission of majority carriers from a trap. Further, the activation energy of the corresponding signal is similar to the one observed in admittance spectroscopy. The DLTS data points in the Arrhenius plot continue the admittance data, suggesting that it is the same signal as the one observed in AS. These results are in accordance with our earlier observations, where a reduction in apparent doping was observed after passivation of the ~200 meV defect,[5,32] and confirm our speculation of the ~200 meV defect being acceptor in nature.

Earlier work has established the presence of deep defects in CISe solar cells, which can be passivated with mild surface chalcogen treatments and buffer layers with high sulfur concentration in the deposition process.[28,32,35] Particularly interesting point is that these buffer layers *i.e.* CdS and Zn(O,S) are deposited *via* chemical bath at low temperature (<85°C), whereas the chalcogen treatment is done at higher temperatures (>300°C).[32] This suggests that the defect is present at or near the surface within few tens of nanometers. To explore this possibility and rule out the properties of buffer layer as a viable cause for the disappearance of defect signature in AS, three post-deposition treatments (PDT) are performed. For the PDTs, KCN etched Cu-rich CISe absorbers were immersed into three separate solutions: ammoniac solution of $ZnSO_4$ (Zn-PDT), ammoniac solution of $CdSO_4$ (Cd-PDT), ammoniac solution of $CH_4N_2S$ (S-PDT), each at 80°C for 10 minutes. These absorbers were made into Schottky device and then AS was performed.

Figure 3a gives the summary of the defect energies obtained after the three PDTs along with the values obtained after CdS and Zn(O,S) buffer deposition. Among the three PDTs, Zn-PDT leads to a complete passivation of the defect, confirmed by the significant reduction in the activation energy of the capacitance step. The energies obtained after the treatment can be attributed to the A2 and A3 acceptor in $CuInSe_2$.[36,37] Whereas, S-PDT results in partial passivation, as it exhibits still the signature of a deep defect in the AS (Figure 3b) with activation energy ~170 meV. For this device particularly, the frequency derivative of AS (Figure 3c), displays broad peaks a feature similar to the un-passivated samples. Also, the main capacitance step in admittance spectra starts to bifurcate into two steps (response 'a' and 'b' in Figure 3b) at low temperatures (<130 K), which might be due to presence of two different defect signatures. For better visualization, the high frequency peak of the curve at 124 K is arbitrarily assigned as primary peak and the other as secondary peak in Figure 3c. Figure 3d shows normalized amplitude of the primary peak plotted vs normalized inflection point (*i.e.* frequency at peak maxima) of the corresponding frequency derivative with the temperature as a parameter. Here, to better resolve the two peaks, the admittance spectra was measured in smaller temperature steps (~3 K). A careful observation of the plot reveals the evolution of the second peak highlighted in red at low temperatures. This establishes the presence of two different defects, which constitute the main step in the admittance spectra of Cu-rich CISe devices. For the untreated absorbers, the presence of similar broad peaks in the $\omega dC/d\omega$ spectra (Figure 2b) indicates, even in that case the capacitance step might be originating from contributions of two defects, one more prominent than the other. Lastly, the AS of Cd-PDT device does not show any reduction of the activation energy of the capacitance step (Figure 3a), confirming that neither $Cd^{2+}$, $SO^{4-}$ or $OH^-$ results in passivation as they are contained in Cd-PDT solution. To summarize Zn treatment leads to a complete passivation of the defects, while S treatment leads to a partial passivation and Cd treatment alone leads to no passivation of the defect. In addition these chemical treatments, ultra-high vacuum (UHV) annealing, which is known to passivate near surface properties of $Cu(In,Ga)Se_2$, also results in passivation of the 200 meV defect (see discussion in SI, Figure S4 and S5).[38-40] Thus, together with this and the PDT results it can be concluded that the 200 meV defect is actually a defect at or near the surface, consisting of two constituents, which can be passivated with proper surface treatment.

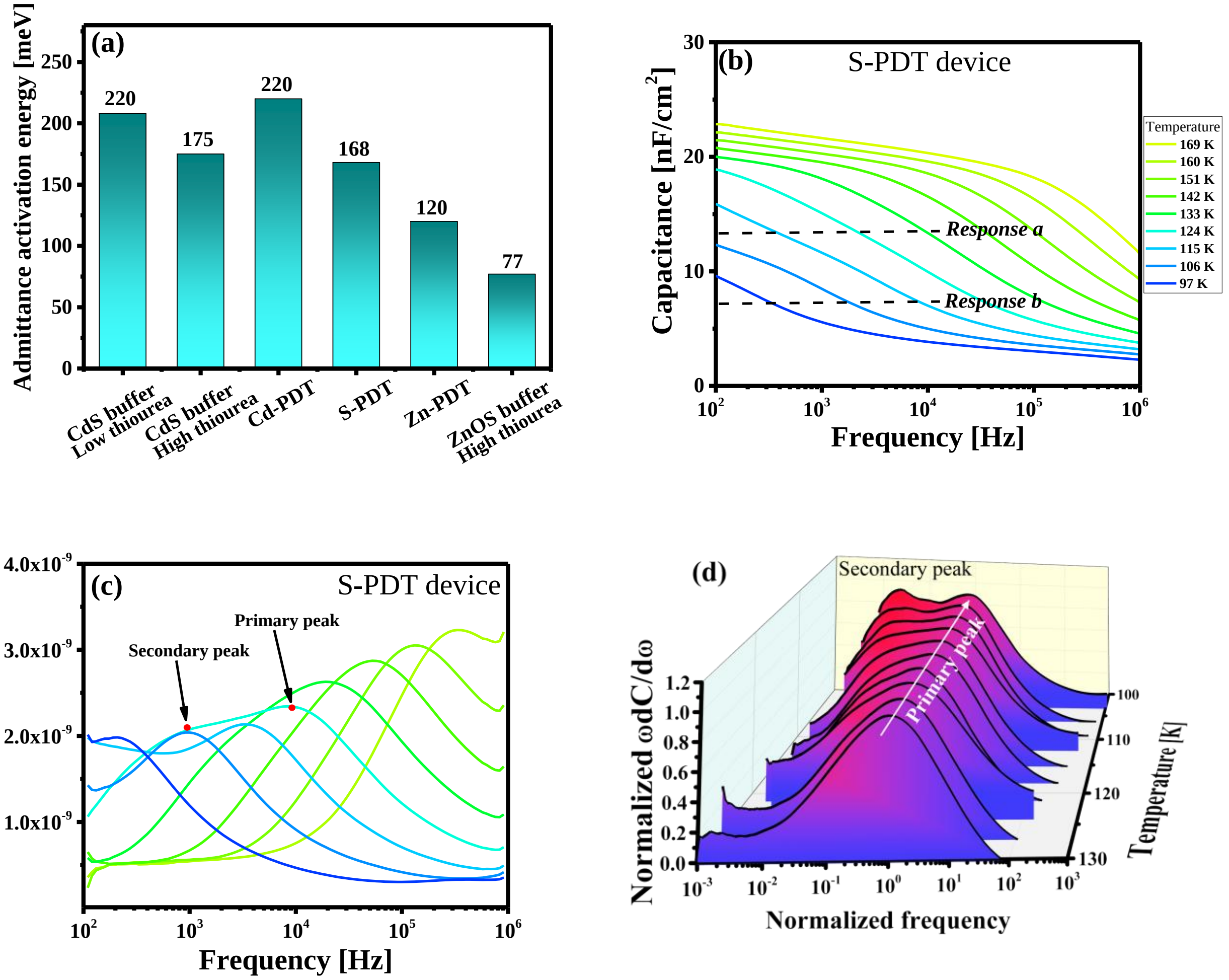

**Figure 3.** (a) Summary of activation energies obtained from Arrhenius plot of the main capacitance step for different PDTs and buffer layers. The bar chart shows the activation energy of the main capacitance step obtained for devices prepared after performing various PDT on the 10% KCN etched absorbers. (b) Admittance spectra of S-

PDT CISe absorber in a Schottky device (c) corresponding ωdC/dω plot, which at 124 K shows double peak structure, the high frequency peak is arbitrarily named primary peak and the low frequency peak as secondary peak. (d) The plot of normalized frequency vs normalized ωdC/dω with respect to frequency. The curve shows the appearance of a secondary peak particularly at low temperature.

To get an estimate of defect density capacitance steps consisting of overlapping defect contributions (see for instance Figure 3d) were fitted as described in [41]. In particular, the defect response from a discrete defect level is extended to Gaussian defect distributions. Here, two Gaussian distributions are used and are fitted simultaneously to the complete temperature and frequency range. A fit describing the two and overlapping capacitance steps of the spectra shown in Figure 3d is shown Figure S6. For untreated sample a defect density ~$2x10^{16}$ $cm^{-3}$ and for S-PDT sample a defect density of ~$4x10^{15}$ $cm^{-3}$ was obtained.

To summarize, the experimental findings: the 200 meV defect is an acceptor defect, has a defect density of around ~$10^{16-17}$ $cm^{-3}$,[32] and is present at or near the surface i.e. it is a sub-surface defect. It is not clear how this defect can lead to the observed large interface $V_{OC}$ loss and to a saturation current activation energy lower than the band gap. In the next section, a numerical model is realized by introducing defects in CISe based on above discussed defect properties with the aim to describe the experimentally observed losses.

**Numerical simulation with sub-surface defects**

The results of the previous section indicate the near-surface and acceptor nature of the defect *i.e.* an acceptor defect present close to or at the absorber/buffer (A/B) interface. Therefore, the defect could represent either a defective layer within the absorber, just below the surface, or a defective interface between the absorber and the buffer. In this section, using numerical modelling, the impact of both, a defective layer and a defective interface on the $V_{OC,in}$ and $V_{oc,ex}$ of the device will be investigated. The models will be assessed to reproduce the experimentally observed characteristics of Cu-rich CISe devices as discussed before: (i) >100 meV interface $V_{OC}$ deficit, (ii) an $E_a$ of the saturation current density lower than the $E_G$ of CISe and (iii) an 'S shape' in the first quadrant at lower temperatures in the I-V curves.

A device model is designed in SCAPS-1D emulating the Cu-rich CISe devices (back contact/CISe/CdS/ZnO/Al:ZnO/front contact). Table S1 records the electrical and optical parameters used in the simulations, which were set constant, taking values from previous measurements,[42-44] and are the same as in our earlier simulations.[33] Further, no conduction band offset at the absorber/buffer (A/B) interface and flat band conditions at the absorber back contact were assumed to keep the model as simple as possible and to avoid convergence problems in SCAPS. Two models were developed. Both models involve deep acceptor defects, since the characteristic defect in Cu-rich CIS is a 200 meV deep acceptor state. The first model comprises a defective layer (often called $p^+$ layer in the literature [45,46]) i.e. a thin layer with high concentration of 220 meV deep acceptor defects, already used in our previous work. There are no deep defects at the interface in this model (Figure 4a).[33] The second model comprises a defective interface, with a significant amount of deep interface acceptor defects above mid-gap at the A/B interface and large electron capture cross-section, to ensure Fermi-level pinning (Figure 4b). The defect level is placed 0.65 eV above the valence band in this model. The defect energy value was chosen to allow for simulating an activation energy for recombination current as close as possible to the experimental values. The values of $F_e$ and $F_h$ give a measure of the density of thermal or photo-generated free carriers in the conduction and valence band, respectively. The high defect density ($N_d$) along with a large electron capture cross-section (reported in table S1) in both models results in strong reduction of electron quasi-Fermi level ($F_e$) and thus, a reduction of the $V_{OC,in}$ near the surface due to Shockley-Read-Hall (SRH) recombination. Consequently, the $V_{OC,ex}$ of the device is reduced. Moreover, in both models the $V_{OC,in}$ is reduced only in a very small region near the A/B interface: ~100 nm for the defective layer and ~50 nm for the defective interface, but is otherwise constant throughout the absorber. This quasi-Fermi level gradient near the surface is observed independent of the carrier mobility. Even in high mobility limit (electron mobility values ~100 cm$^2$/V-s), the $V_{OC,in}$ is reduced near the surface in the CISe device. A $V_{OC,in}$ measurement by photoluminescence (PL) reflects the (nearly constant) maximum $V_{OC,in}$ in the bulk of the absorber, as the PL intensity increases exponentially with the $V_{OC,in}$. The $V_{OC,ex}$ is the difference between the majority quasi-Fermi levels on either side. Since there is only a negligible gradient in the hole quasi-Fermi level, the $V_{OC,ex}$ is given by the $V_{OC,in}$ at the absorber buffer interface. Hence, it is established that both models result in deficit between the measured $V_{OC,in}$ and the $V_{OC,ex}$, as depicted in Figure 4.

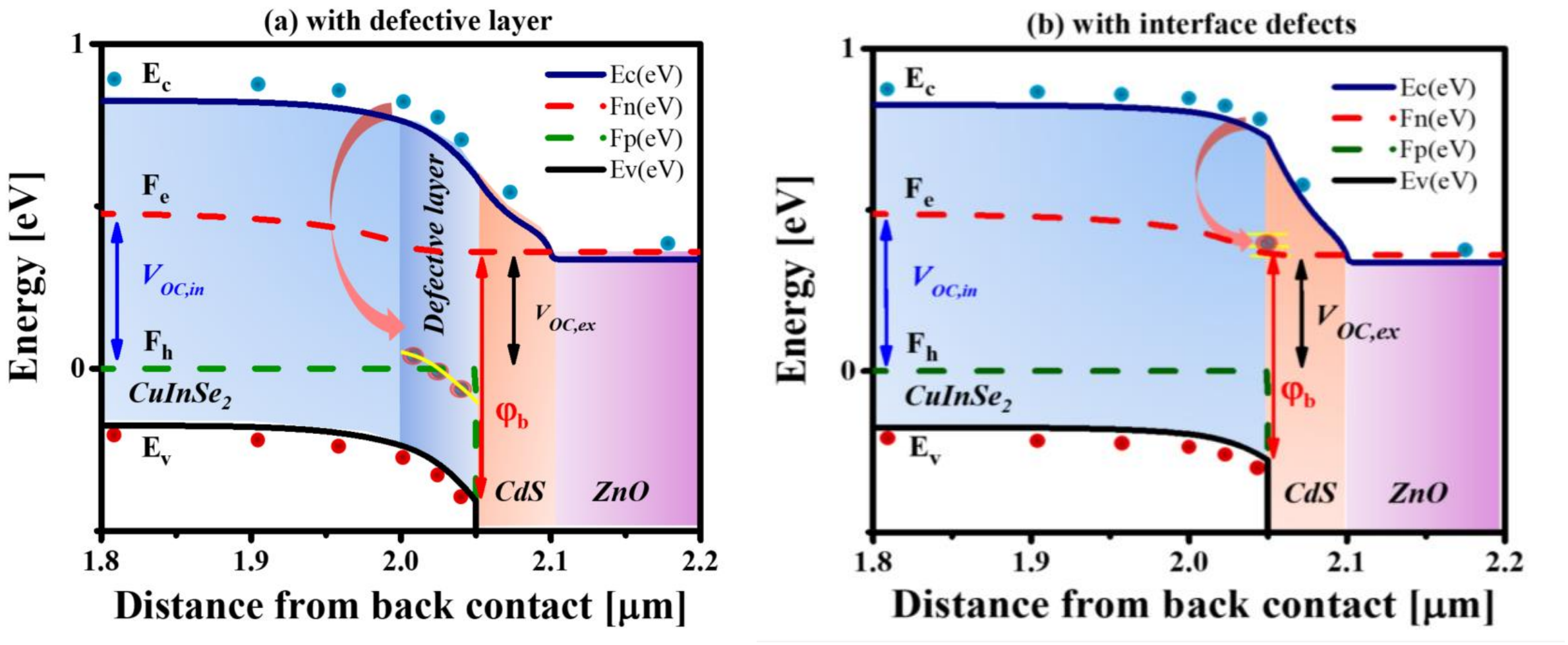

**Figure 4.** Simulated band diagram of the device at open-circuit ($V_{oc}$) voltage with (a) defective layer and (b) interface defects. The maximum quasi-Fermi level splitting in the device is labelled as $V_{OC,in}$, whereas the $V_{OC,ex}$ values are represented as the difference between the hole quasi-Fermi level at the back contact and electron fermi

level at front contact. The yellow line shows the defect levels with high concentration in the device structure and $\varphi_b$ is the hole barrier at the interface.

As demonstrated in Figure 4, both models are capable of reproducing the experimentally observed $V_{OC,in}$ and $V_{OC,ex}$, and hence, the interface $V_{OC}$ deficit. However, the validation of either model as the appropriate description for Cu-rich CISe devices requires also fulfillment of criteria (ii) and (iii). All Cu-rich chalcopyrite devices are characterized by a saturation current, strongly dominated by interface recombination. This is indicated by the $E_a$ obtained from extrapolating $V_{OC,ex}$ *vs* temperature is always lower than the $E_G$.[10,19] As shown before, the Cu-rich CISe devices presented here also suffer from the same issue. Two possible explanations for an activation energy of the saturation current $E_a$ lower than the bandgap are established in the literature: a cliff at the absorber buffer interface, *i.e.* conduction band minimum of CdS lower than that of CISe, or Fermi level pinning at this interface.[19,47] Thus, a straightforward origin of interface recombination could be an unfavorable band offset, *i.e.* a cliff at the interface. However, CdS is a perfectly suited buffer for Cu-poor $Cu(InGa)Se_2$ absorbers, which have a higher conduction band minimum than pure $CuInSe_2$. There is no indication that the band edges of Cu-rich $CuInSe_2$ are different from those of Cu-poor material. Furthermore, the photoelectron study by Morkel *et. al.* reports a conduction band minimum of CdS aligned with the one of CISe, eliminating unfavorable band offset as the possible cause for interface recombination.[48] The other possible scenario could be the presence of a high concentration of defects ($N_{IF}$) at the CISe/CdS interface, which pins the electron Fermi-level at the interface. In order to have a working solar cell like in Figure 1a, the pinning position must be above the middle of $E_G$ to obtain a decent $V_{OC,ex}$. Thereby the electron concentration at the interface remains significantly higher than the hole concentration. Thus, making the interface recombination dependent on the interface hole concentration ($p_{IF}$) and the hole surface recombination velocity ($S_p$) i.e. R ≈ $p_{IF}*S_p$.[19] The reverse saturation current density ($J_0$) then is given by:[19]

$$J_0 = qN_{v,a}S_p \exp\left\{-\frac{\varphi_b}{kT}\right\} \quad (1)$$

Where, $N_{v,a}$ is the effective valence band density of states in the absorber and q is the elementary charge, $\varphi_b$ is the equilibrium hole barrier at the interface and is equal to the energy difference between the position of electron fermi level ($F_e$) and the valence band edge ($E_v$) *i.e.* $\varphi_b = F_e - E_v$. Equation (1) is true if the recombination current is dominated by interface recombination, *i.e.* in the case of a significant $S_p$. This is more likely for a negatively charged interface, *i.e.* with a high density of acceptor states. However, it is not necessarily the case that the pinning defect and the recombination defect are the same, although this is what we assume in our simulation. From equation 1 it is evident that in case of Fermi-level pinning the $E_a$ of the saturation current should be $\varphi_b$, which is lower than the $E_G$. Consequently, the open-circuit voltage is given as follows:[19,47]

$$V_{OC,ex} = \frac{\varphi_b}{q} - \frac{nkT}{q}\ln\left(\frac{qN_{v,a}S_{p0}}{J_{ph}}\right) \quad (2)$$

Where $J_{ph}$ is the photogenerated current, $n$ is the diode ideality factor. Thus, $V_{OC,ex}$ is dominated by $\varphi_b$. One should note that in a good device without interface recombination, the $V_{OC,ex}$ at 0K is equal to the bandgap of the absorber. Thus, in case of Cu-rich CISe device with spike-type band alignment, Fermi-level pinning could explain an $E_a$ value smaller than the $E_G$, namely $\varphi_b$ obtained from $V_{OC,ex}$ *vs* temperature plot (assuming $n$, $S_p$ and $J_{ph}$ are not or only weakly temperature dependent). We will therefore investigate further predictions from this model in the following.

For conceiving the appropriate defect model for CISe by numerical simulations, the device performance as displayed in Figure 1 will be simulated. Figure 5a shows the simulated $V_{oc}$ values at different temperatures obtained from the two models with defects at or near the interface and for a reference model without any near interface defects. The simulations go down to 210 K, at lower temperatures the numerical calculations would no longer converge. Remarkably, not only the model with electron Fermi-level pinning but, also the model with a defective layer leads to an $E_a$ of the saturation current less than the absorber $E_G$. It should be noted, that the main recombination in the device with defective layer occurs in that defective layer and not at the interface [Figure S7a]. The $E_a$ values obtained with this model are slightly higher than experimental values. Even a considerable increase in defect concentration does not result in an $E_a$ value below 0.81 eV [Figure S8a]. Also, above a certain value the defect concentration starts to limit the short-circuit current ($J_{sc}$) of the device meaning that the model would be no longer realistic (Figure S8b). Thus, we kept the defect density at a value that still gives a realistic short circuit current density, although it cannot fully describe the reduction of $E_a$. This effect is also not unexpected, since we kept the model rather simple, to allow for temperature dependent simulations.

Thus, both models are capable of introducing a recombination pathway with an $E_a$ lower than the $E_G$. Another important observation comes from hole barrier simulation at different temperatures (Figure 5b). Neither of the two model results in a temperature independent hole barrier ($\varphi_b$). However, the $\varphi_b$ exhibits a weak temperature dependence in the device with interface defect model, and the extrapolation of $\varphi_b$ to 0 K equals the $E_a$ obtained from $V_{OC,ex}$ measurements. This indicates that the simple model of Fermi-level pinning in eq. 1 is only an approximation, and $E_a$ should be identified as $\varphi_b$ at 0 K as the $\varphi_b$ itself is weakly temperature dependent. It is noteworthy, that the $N_{IF}$ used here was $10^{12}$ $cm^{-2}$ and even $N_{IF}$ of $10^{14}$ $cm^{-2}$ results in a weakly temperature dependent $\varphi_b$. Even in the latter case $E_a$ is not equal to the $\varphi_b$ at 300 K. On the contrary, in the device with the defective layer the $\varphi_b$ extrapolates to $E_G$ at 0 K and is strongly temperature dependent. In this case $E_a$ of the recombination current is not determined by the hole barrier. Yet, a strongly defective layer can also lead to activation energies lower than $E_G$ - without Fermi-level pinning, and without a cliff in the conduction band alignment.

Finally, we test the model on criterion (iii), *i.e.* the 'S shape' in the first quadrant exhibited by Cu-rich CISe devices at lower temperatures. Figure 5c shows the I-V curves at low temperatures simulated for a device with a defective layer and a device with defective interface. For the first model 'S shape' in I-V at low temperatures in the first quadrant is observed, as established previously due to the presence of $p^+$ layer (defective layer) near the interface.[33] On the contrary, the presence of Fermi-level pinning at interface leads to an 'S shape' in the fourth quadrant, signifying that it rather acts as a barrier for extraction of photogenerated carriers. Thus, the I-V-T behavior of the device is best described by the model with a defective layer.

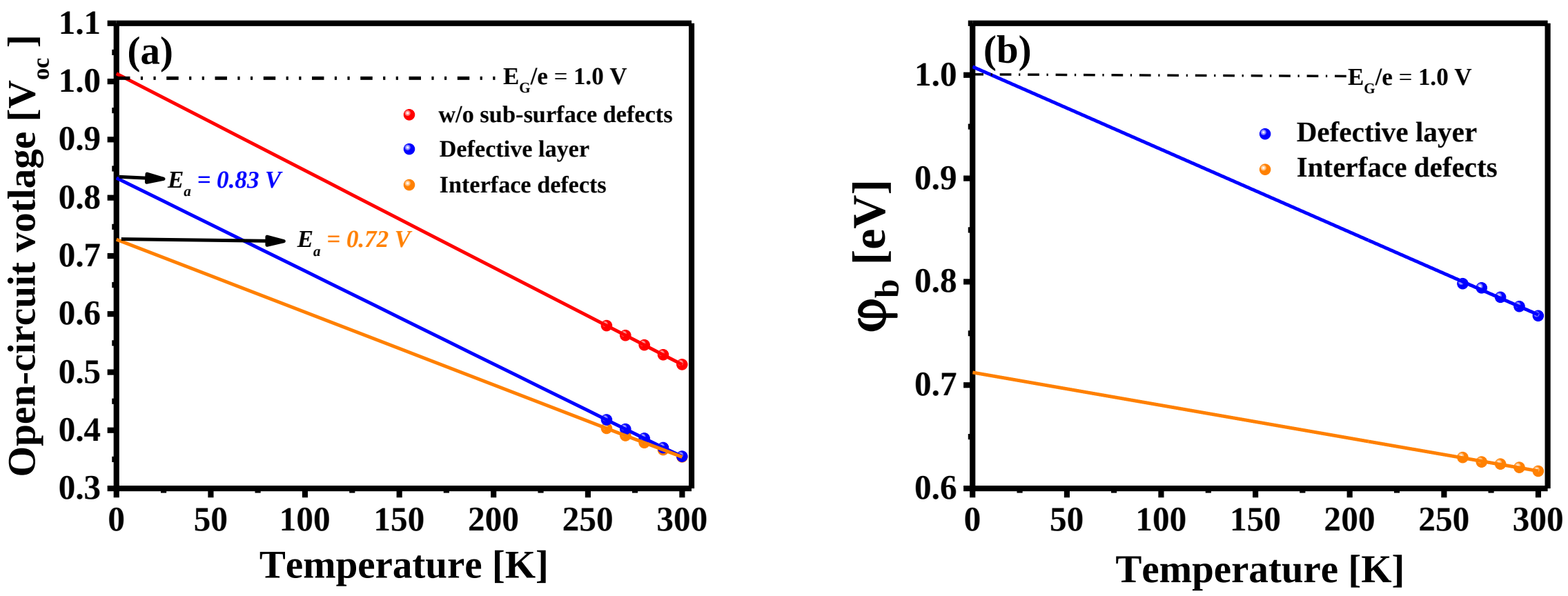

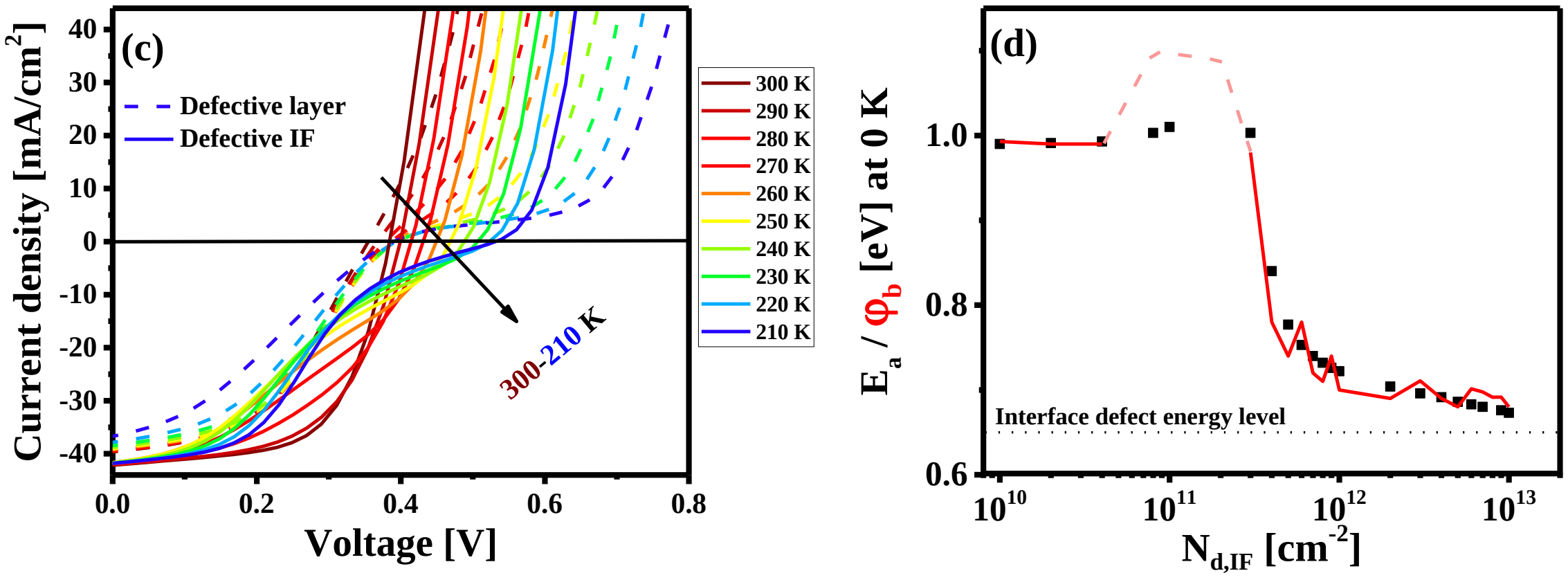

**Figure 5.** (a) Simulated open-circuit voltage ($V_{oc}$) values of the device with defective layer and of the device with interface defects. (b) The hole barrier as the function of temperature and its extrapolation to 0 K. (c) Simulated I-V curve at different temperatures of devices with defective layer and with defective interface. The former results in 'S shape' in first quadrant (solid lines), whereas, later results in 'S shape' in fourth quadrant (dashed lines). (d) Activation energy ($E_a$) and the hole barrier ($\varphi_b$) at 0 K for the device with interface defects as a function of interface defect density. The $E_a$ is obtained from $V_{OC}$ vs temperature curve and the $\varphi_b$ is obtained from extrapolation of hole barrier to 0 K. The graph clearly shows a direct correlation of the activation energy with hole barrier height. Both quantities approach the energy of the defect at high defect concentrations. We believe that the two points with $\varphi_b$ larger than the $E_G$ are a numerical artefact.

It is established that the model with a defective layer explains, to a good extent, the experimentally observed Cu-rich CISe device characteristics. At this point, it is worth summarizing a few points regarding both models. Both models lead to a significant interface $V_{OC}$ deficit in the device and an $E_a < E_G$. The exact value of both $V_{OC}$ deficit and $E_a$ depend on the defect properties such as defect energy, density and capture cross section. However, the exact mechanism in the two cases is different: in the defective layer model, the main recombination is in the SCR close to the surface and does not result in Fermi-level pinning [Figure 5b]. In the contrary, the CISe/CdS interface is the location of the main recombination channel in the defective interface model, and leads to a weak Fermi-level pinning as evident from Figure 5b where the $\varphi_b$ changes only weakly with temperature. The $E_a$ is given by the value of $\varphi_b$ at 0 K. Figure 5d shows simulated $E_a$ and $\varphi_b$ at 0 K (obtained by extrapolating simulated hole barrier values to 0 K), as in Figure 5 a and b as a function of interface defect density ($N_{d,IF}$). It is clear that in a certain range by varying the defect density one can have $E_a$ anywhere between the $E_G$ and the defect position in the interface $E_G$. Further, there is a one-to-one correlation between $E_a$ and $\varphi_b$ at 0 K.

Even though the models presented here might be not fully accurate, as they do not include many factors such as surface $E_G$ widening or band offsets between absorber and buffer. Still, the models do a good job of reproducing the main experimental characteristics of Cu-rich CISe devices that indicate a problematic interface, and provide a suitable explanation. Out of the two models, the defective p+ layer explains better the observed I-V behavior at low temperatures. In addition, the simulations demonstrate that the commonly used model of eq. 1 is only an approximation, yet a useful one. Furthermore, we showed that the most critical parameters indicating interface recombination *i.e.* a significant difference between $V_{OC,in}$ and $V_{OC,ex}$, and an $E_a$ of saturation current lower than the $E_G$ can be reproduced by a model that contains neither a reduced interface bandgap, nor Fermi level pinning.

Moreover, these models, though applied and developed for Cu-rich CISe device, are equally applicable to any other device as well. Particularly, for heterojunction devices, which have optimum band-offset with the hole and electron transport layer, but are still dominated by interface recombination. Other than the conventional Fermi-level pinning, the interface recombination signature in this case could alternatively originate from the defective surface layer. The results of the simulations also demonstrate a way to differentiate between defective surface and defective interface. In both cases, the temperature dependent $V_{OC,ex}$ measurements will yield an $E_a$ for saturation current lower than the $E_G$. However, the two models can be distinguished looking at the I-V curves. While a defective interface results in 'S shape' in the first quadrant, the defective surface results in 'S shape' in the fourth quadrant. Once the root cause *i.e.* the presence of either defective interface or defective surface is identified, a dedicated passivation strategy can be used to improve the device performance.

**Conclusions**

In summary, CISe absorbers grown under Cu-excess and Cu-deficient conditions although characterized by similar $V_{OC,in}$ possess different $V_{OC,ex}$ in the device due to presence of near surface defects. DLTS measurements revealed these defects are acceptor in nature. The presence of these acceptor defects in Cu-rich device lead to significant interface $V_{OC}$ deficit leading to lower efficiency and electronic barriers in device structure, which is not observed in Cu-poor device. To elucidate the root origin of interface $V_{OC}$ deficit, we have demonstrated two comprehensive models for Cu-rich CISe solar cells, which can be applied to other hetero-structure solar cells as well. These models comprise either a near interface layer or the interface itself with a high concentration of deep acceptor defects. The drift and diffusion simulations have demonstrated that both models are capable of reproducing electrical characteristics of Cu-rich CISe devices, in particular reduced $V_{OC,ex}$ compared to $V_{OC,in}$. The reduction emanates due to deep traps at or near the surface which lead to strong non-radiative recombinations in the region near the surface and dominate the $V_{OC,in}$ near the surface. As a consequence, the quasi-Fermi level splitting decreases rather abruptly near the surface resulting in a reduced $V_{OC,ex}$, thus resulting in an interface $V_{OC}$ deficit in the device. In cases as such, the information regarding the gradient $V_{OC,in}$, is not accessible from PL measurements. However, we have demonstrated that the presence of both a defective surface and a defective interface could be confirmed by temperature dependent $V_{OC,ex}$ measurements. In both cases the activation energy of the saturation current density obtained by temperature dependent $V_{OC,ex}$ measurements is lower than the bulk $E_G$ of the absorber. Furthermore, we show that the presence of either defective layer or defective interface in a device predict an activation energy of the saturation current lower than the $E_G$ and can be differentiated through I-V measurements particularly at temperatures below 300 K. While the defective surface leads to a 'S shape' in the first quadrant of the I-V curve as a signature of a barrier for injected carries, as observed experimentally, the defective interface leads to a 'S shape' in the fourth as a signature of a barrier to photogenerated carriers.

Particularly for Cu-rich CISe solar cells together with AS and DLTS spectroscopy, a model is developed, which correlates the interface $V_{OC}$ deficit to the presence of acceptor defect in Cu-rich CISe absorbers. A comparison of AS of absorbers etched with aqueous KCN and aqueous Bromine solutions revealed the defect to be an intrinsic part of Cu-rich devices originating from the etching of secondary phases, independent of the etchant. DLTS confirms that this defect is an acceptor defect. Analysis of several PDTs on the CISe absorbers demonstrated that the usual broad AS defect signature is produced by the response from two defect levels close to each other.

As a general point of view, calibrated PL measurements provide information regarding the ratio of non-radiative to radiative recombination in the bulk of the absorber. However, in these measurements near surface properties could be overlooked. To account for these there is the need of complimentary techniques such as temperature dependent I-V measurements to characterize the device and assign recombination channels in the device. We have provided two universal models which can also be applied to others photovoltaic technologies to explain and understand the cause of interface $V_{OC}$ deficit in the case where the band alignment does not impose a cliff situation.

**Device preparation and characterization methods**

For the experiments, we used polycrystalline CISe thin films grown on molybdenum coated soda lime glass in a 1-stage process. Comprehensive details of the deposition process can be found in our previous report.[32] For investigating the impact of Zn, Cd, S post deposition treatments, the CISe absorbers were etched with 10 % KCN solution for 5 minutes to remove the $Cu_{2-x}Se$ secondary phase. These were then immersed in three separate solutions; $3CdSO_4.8H_2O$ (0.1M) in $NH_4OH$ (2M), $ZnSO_4.7H_2O$ (0.1M) in $NH_4OH$ (2M), and $CH_4N_2S$ (0.4M) in $NH_4OH$ (2M) at 84 °C for 15 minutes, all freshly prepared. For bromine treatment, the un-etched absorbers were immersed in aqueous $Br_2$ (0.01M) plus potassium bromide (0.3M) solution for 1 minute. The treatment schematic can be seen in Figure S9.

The treated absorbers were further processed into two device configurations for characterization: “Schottky device” [$CuInSe_2$ with aluminum dots] and “Solar cell” [$CuInSe_2$ coated with CdS followed by zinc-oxide, aluminum doped zinc-oxide, and nickel aluminum grids]. A standard Xenon short-arc lamp AAA solar simulator calibrated with a reference Si solar cell, with an IV source-measure-unit was used to measure the I-V of the devices. To perform low temperature electrical characterization (AS, DLTS and I-V-T) the devices were mounted inside a closed-cycle cryostat under vacuum below $4x10^{-3}$ mbar. A cold mirror halogen lamp adjusted to an intensity of ~100 $mW/cm^2$ was used to illuminate the device for I-V-T measurements. An inductance, capacitance, and resistance (LCR) meter was used to measure the admittance of the sample. In the setup a controlled small-signal ac voltage pulse of 30 mV rms with frequency from f=20 Hz to 2 MHz was applied. In order to ensure accurate determination of device temperature during all the characterization, a Si-diode sensor glued onto an identical glass substrate was placed beside the solar cell. The numerical simulations were executed using SCAPS1-D software developed at the department of Electronics and Information Systems (ELIS) of the University of Gent, Belgium.[49]

**Acknowledgements**

This research was funded in whole, or in part, by the Luxembourg National Research Fund (FNR), grant reference [PRIDE 15/10935404/MASSENA], [SUNSPOT 11244141], [SURPASS 11341159/SURPASS] [CURI-K C14/MS/8267152 CURI-K], [CORRKEST C15/MS/10386094/CORRKEST] project. For the purpose of open access, the author has applied a Creative Commons Attributions 4.0 International (CC BY 4.0) license to any Author Accepted Manuscript version arising from this submission.

We thank Prof. Małgorzata Igalson, Dr. Alex Redinger and Dr. Sudhanshu Shukla for their valuable discussions and feedback on the present work. We are also thankful to Dr. Marc Burgelman and his team at the University of Ghent, Belgium, for providing SCAPS-1D simulation software.

**Conflict of Interest**

The authors declare that they have no financial/commercial conflict of interest.

**Availability of data**: The data of Fig 1a, 1c, 2a, 3b, and 3d along with the SCAPS definition file is available at 10.5281/zenodo.4643534.

# SI: Near surface defects: Cause of deficit between internal and external open-circuit voltage in solar cells

Mohit Sood*[1], Aleksander Urbaniak[2], Christian Kameni Boumenou[1], Thomas Weiss[1], Hossam Elanzeery[1], Finn Babbe[1,3], Florian Werner[1,4], Michele Melchiorre[1], Susanne Siebentritt[1]

(E-mail address: mohit.sood@uni.lu)

[1]*Department of Physics and Materials Science, University of Luxembourg, Belvaux, L-4422, Luxembourg*

[2]*Faculty of Physics, Warsaw University of Technology, Koszykowa 79, Warszawa 00-662, Poland*

[3]*Chemical Sciences Division, Joint Center for Artificial Photosynthesis, Lawrence Berkeley National Laboratory, Berkeley, USA*

[4]*Hydrosat, 9, rue du Laboratoire, L-1911 Luxembourg*

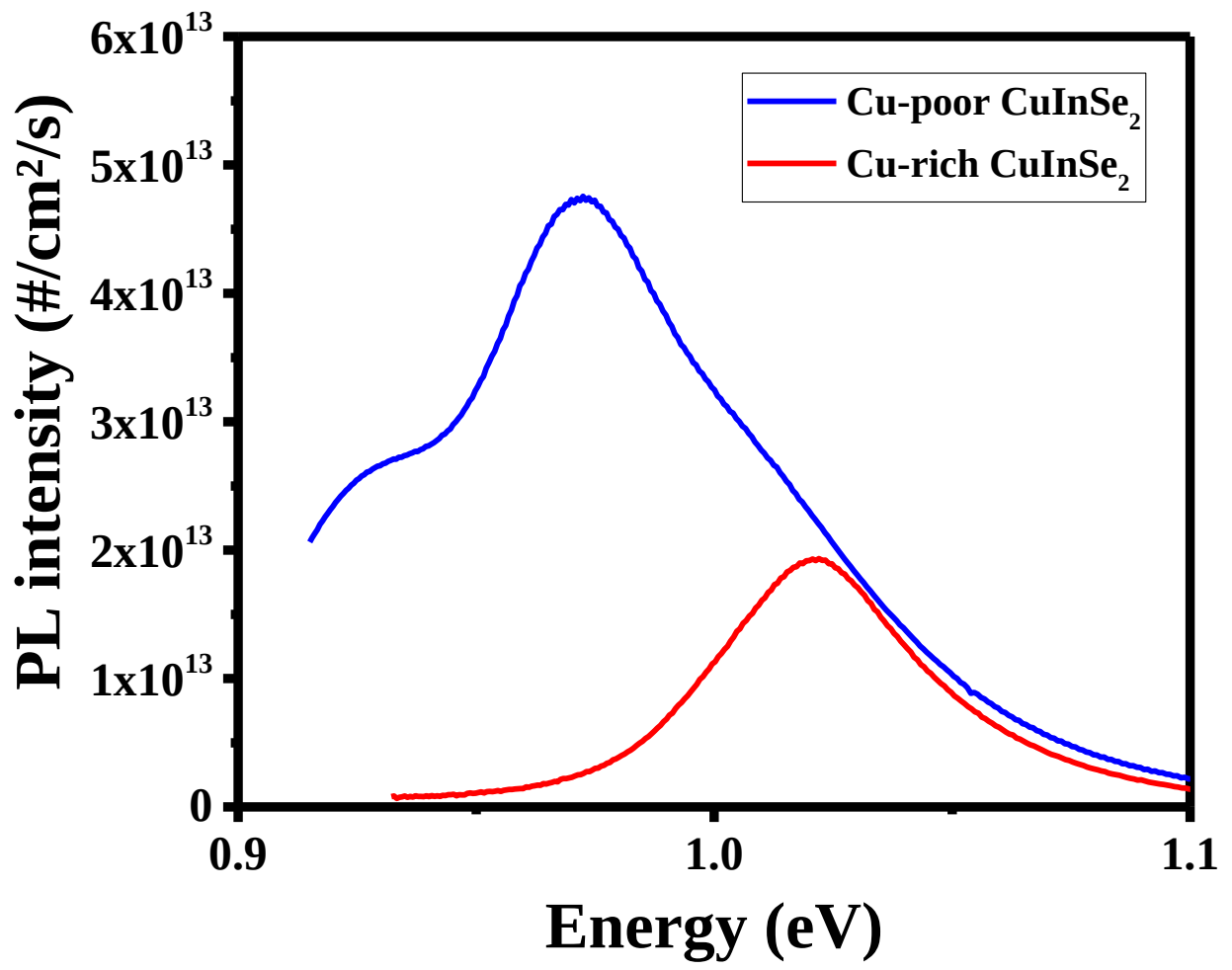

***Figure S1***. Exemplary measured calibrated PL spectra of Cu-rich and Cu-poor $CuInSe_2$ absorbers covered with CdS buffer layer.

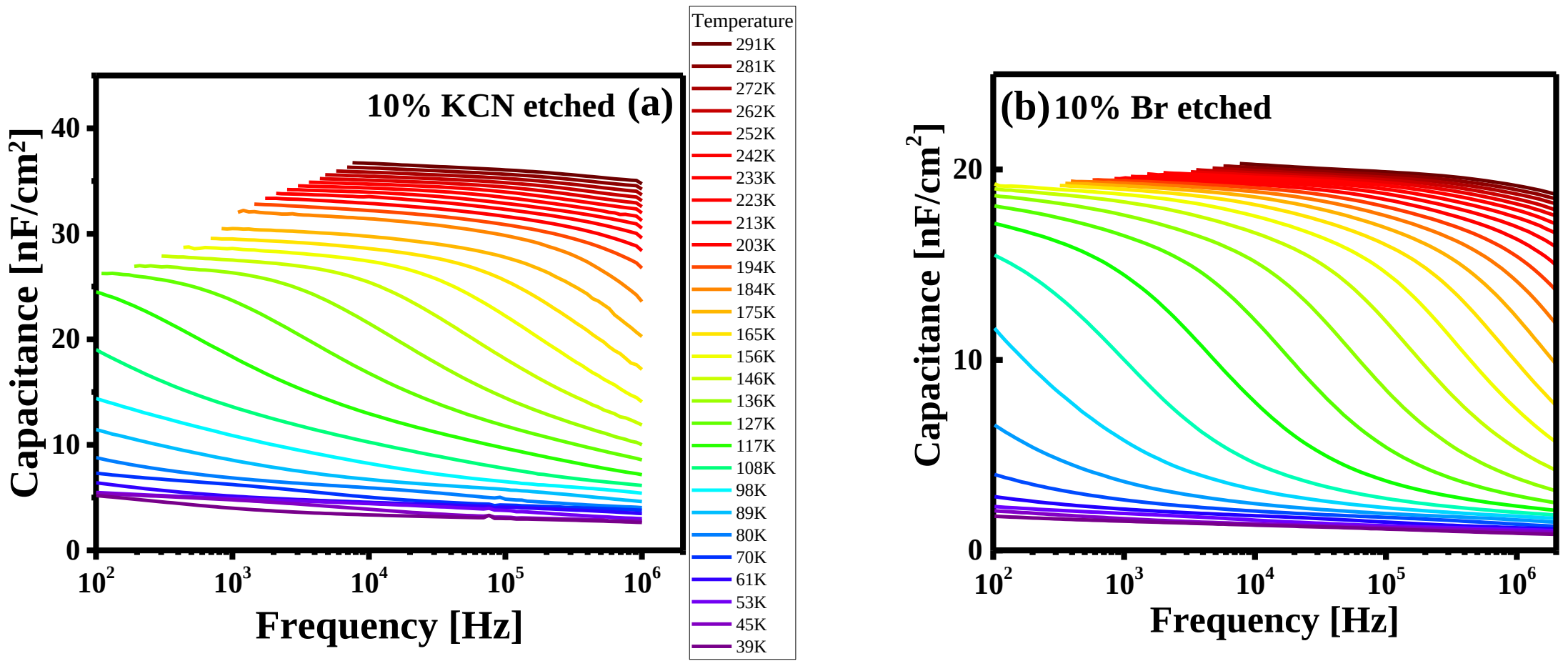

***Figure S2***. Admittance spectra of $CuInSe_2$ device with schottky contact with (a) 10% KCN etching and (b) 0.01M aqueous Br etching.

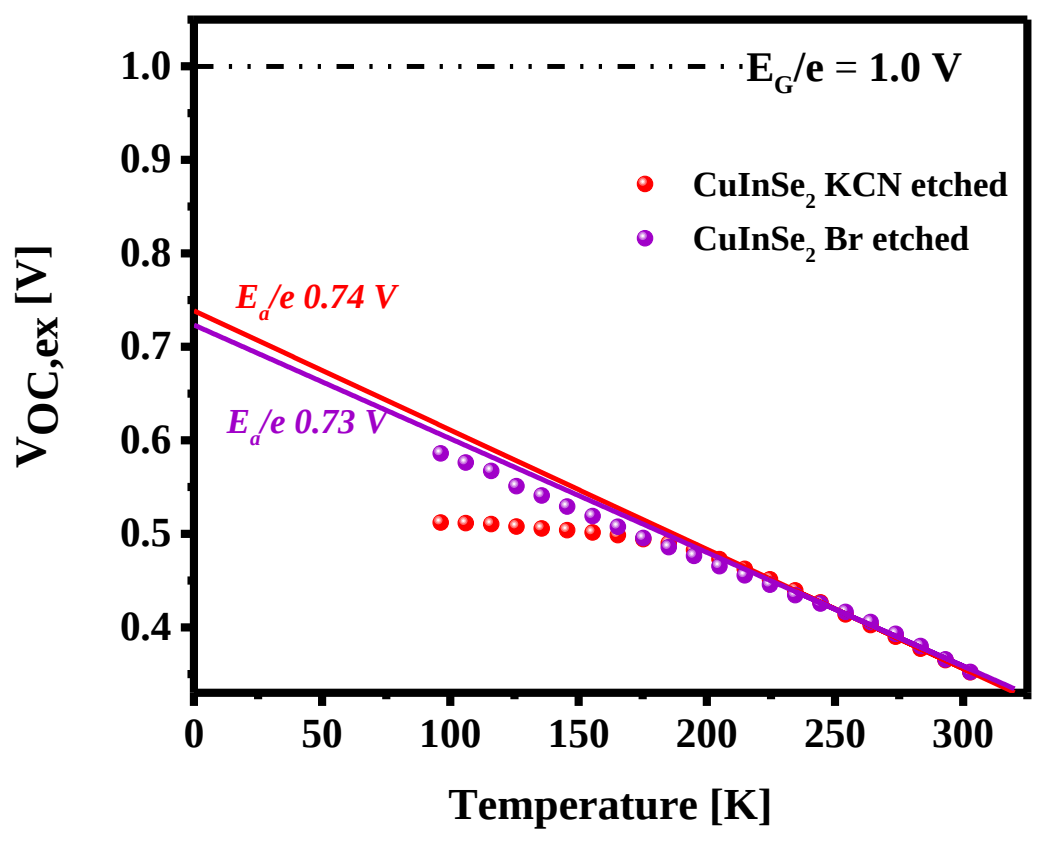

***Figure S3.*** (a) $V_{OC,ex}$ measurements of 10% KCN and 0.01M Br solution etched $CuInSe_2$ solar cells.

## Passivation of near surface defect by UHV annealing

The UHV annealing at 280 °C for 30mins at a base pressure < 2 x 10 $^{-9}$ mbar, was performed on a Cu-rich CISe absorber to assess its ability to passivate sub-surface defects via measuring its impact on the ADM spectra, particularly on the deep defect signature. After UHV annealing the absorber along with a reference sample were finished into solar cell using the baseline process. Figure S3a displays the ADM spectra of reference device, where the highlighted capacitance step in the middle is the commonly observed ~200meV defect. The signature capacitance step disappears in the device prepared with UHV annealed absorber Figure S3b, thus leaving only a step with activation energy of ~86meV. Figure S3c shows the inflection point of capacitance data plotted in Arrhenius plot together with results presented earlier for a better comparison. Apart from UHV annealed device's data points, all other data points scatter very close to the same line and therefore very likely originates from the same defect signature. This indicates the passivation of the 200meV defect at least to the detection limits of ADM. The passivation is further supported by the activation energy $E_a$ of the dominant recombination channel obtained by $V_{OC,ex}$ extrapolation of the device.

As stated before the $V_{OC,ex}$ extrapolation at 0K of the devices dominated by sub-surface defects does not go to the band gap but rather to a value less than the bandgap determined from inflection point in external quantum efficiency measurements. For the UHV annealed device $V_{OC,ex}$ extrapolation goes almost to the bandgap (Figure S4). Therefore, from ADM spectroscopy and $V_{OC,ex}$ extrapolation we conclude that the UHV annealing is an alternate passivation method for the 200meV defect.

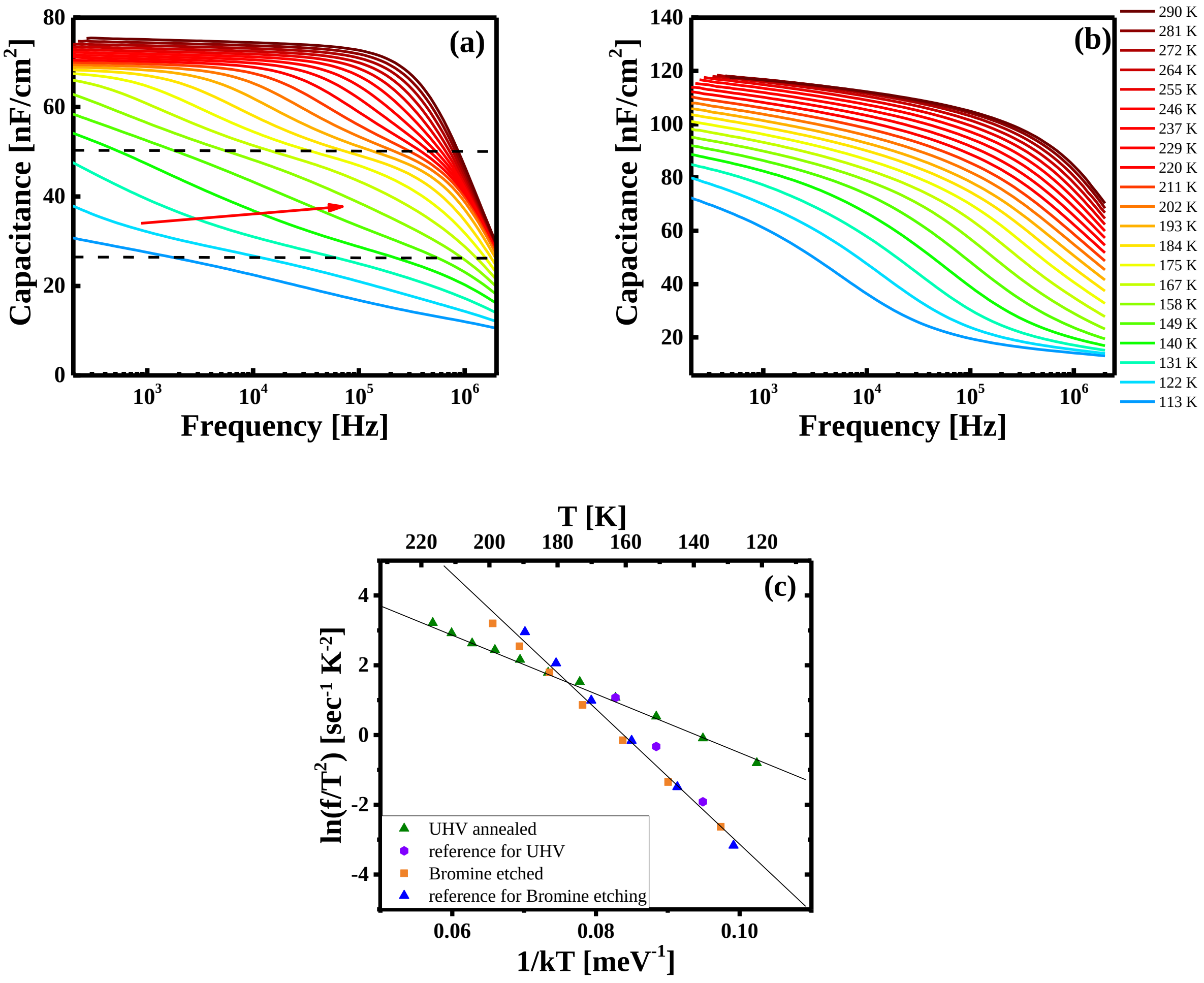

***Figure S4***. Admittance spectra of CISe solar cell prepared from (a) 10% KCN etched absorber (b) 10% KCN etched UHV annealed absorber. (c) Arrhenius plot of KCN etched and UHV annealed solar cell along with KCN etched and bromine etched Schottky device.

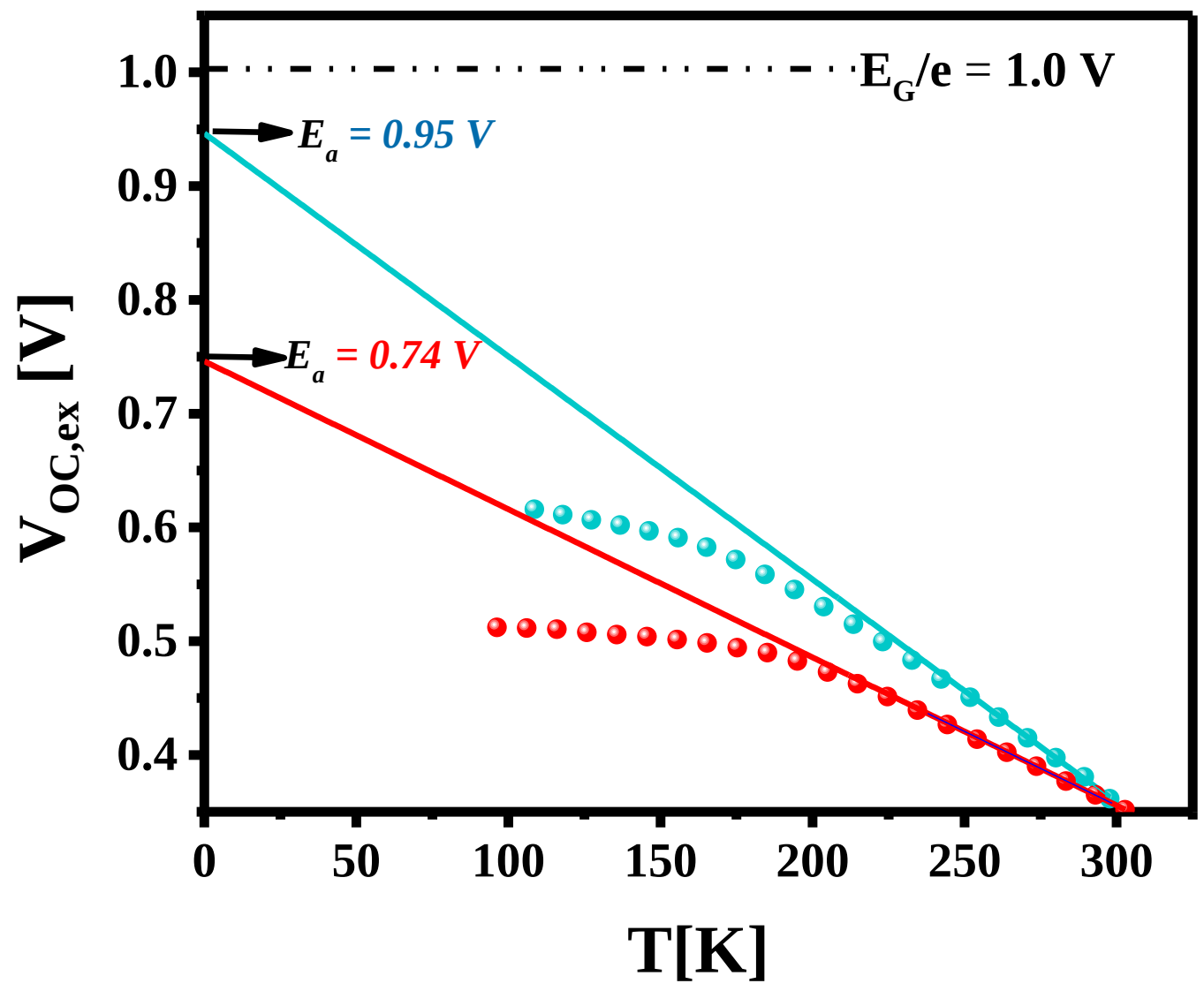

***Figure S5.*** External open-circuit voltage ($V_{OC,ex}$) measurements of $CuInSe_2$ device prepared with 10 % KCN etched absorber and UHV annealed 10% KCN etched absorber.

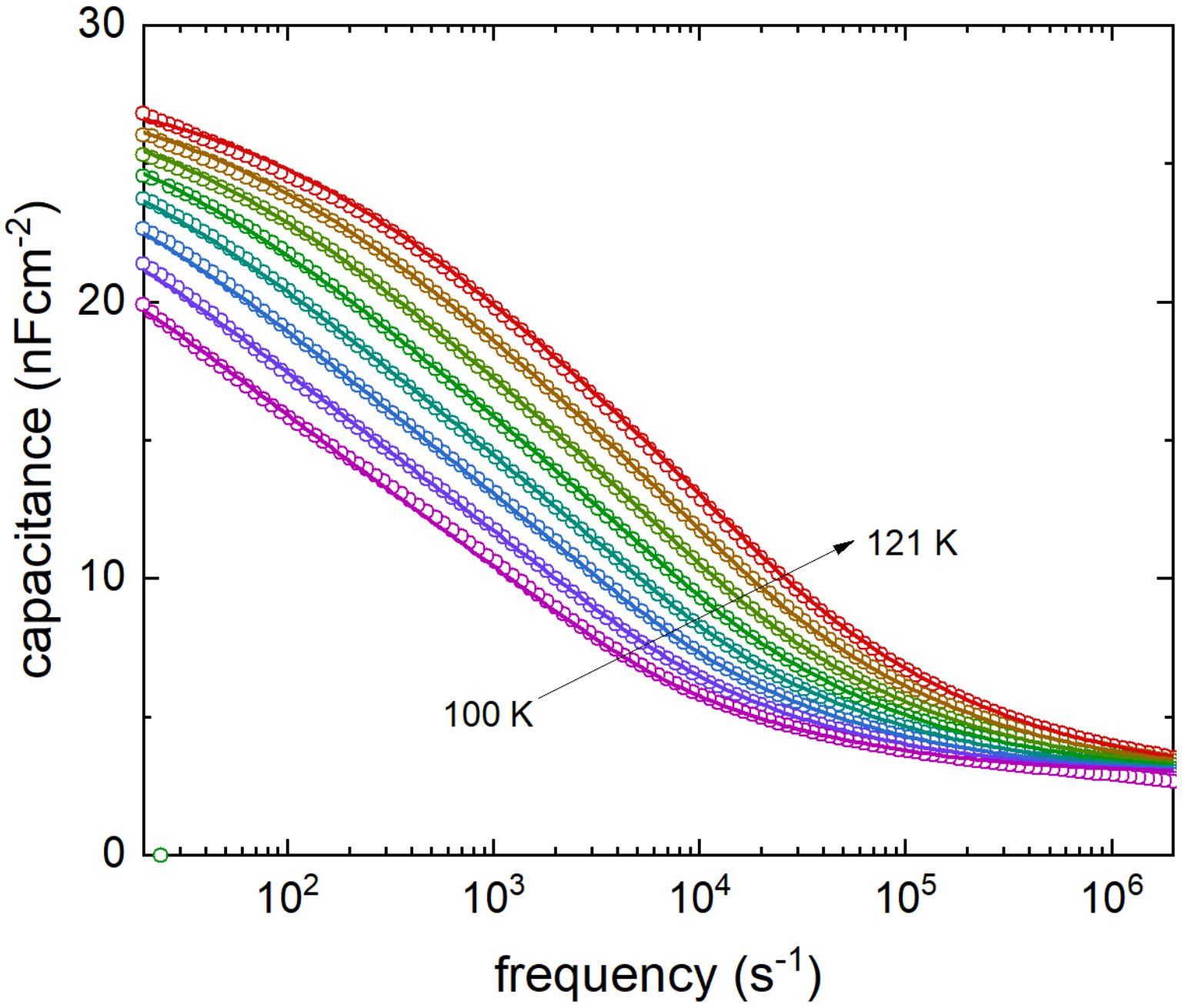

***Figure S6.*** Experimental data (open symbols) and the corresponding fit (solid lines) for the double capacitance step of the sample shown in Figure 3d.

***Table S1.*** *Simulation parameters used to in this work. For achieving a value of $V_{OC,in}$ comparable to as observed in optical measurements, a deep defect level at 300 meV is introduced in the CISe absorber layer.*

| Parameter | $CuInSe_2$ | $p^+$ $CuInSe_2$ | CdS | ZnO/AZO | IF defects $CuInSe_2$/CdS |
|---|---|---|---|---|---|
| **Thickness (µm)** | 2.5 | 0.05 | 0.050 | 0.350 | - |
| **Band gap(eV)** | 1.0 | 1.0 | 2.40 | 3.45 | - |
| **Dielectric permittivity (relative)** | 13.6 | 13.6 | 10 | 10 | |
| **Electron affinity(eV)** | 4.6 | 4.6 | 4.5 | 4.6 | - |
| **Electron mobility($cm^2$/Vs)** | 20 | 20 | 50 | 50 | - |
| **Hole mobility($cm^2$/Vs)** | 10 | 10 | 20 | 20 | - |
| **Doping(1/$cm^3$)** | $1x10^{16}$ | $1x10^{16}$ | $1x10^{16-17}$ | $1x10^{17-19}$ | - |
| **Defect density(1/$cm^3$) Single acceptor from $CuInSe_2$ valence band** | $1x10^{16}$ 300meV | $1x10^{16}$ 300meV & $5x10^{16}$ 220meV | - | - | $1x10^{16}$ 650meV |
| **Capture cross-section electrons ($cm^{-2}$)** | $1x10^{-15}$ | $1x10^{-12}$ for 200meV | - | - | $1x10^{-12}$ |
| **Capture cross-section holes ($cm^{-2}$)** | $1x10^{-15}$ | $1x10^{-13}$ for 200meV | - | - | $3x10^{-16}$ |

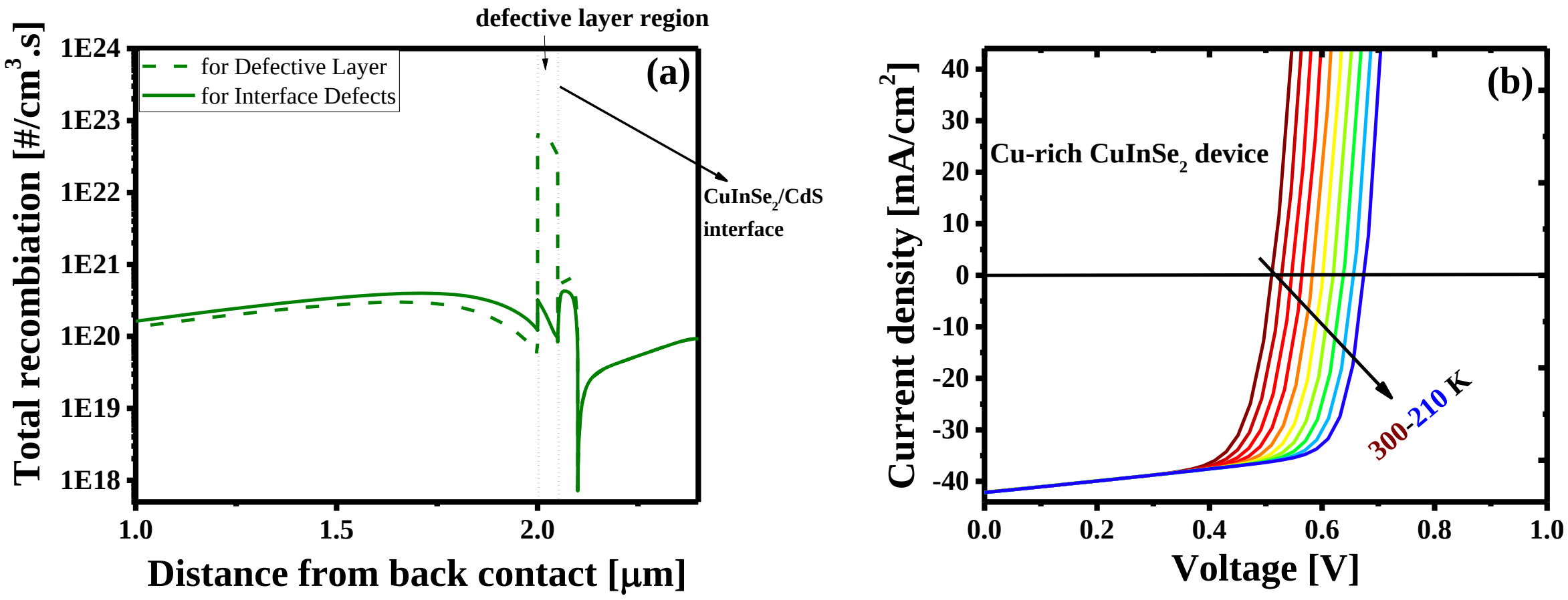

***Figure S7.*** (a) Generation (blue) and recombination (red and green) profiles as a function of distance from the back contact at $V_{OC,ex}$ for simulated devices with a defective layer and with interface defects. For the defective layer model (dashed line) dominant recombination appear to occur near the surface of the absorber, whereas in defective interface model (solid line) they occur at the interface. (b) Simulated I-V curve of a reference device with no defective layer no defective interface. The curves show no rollover or ‘S’ shape.

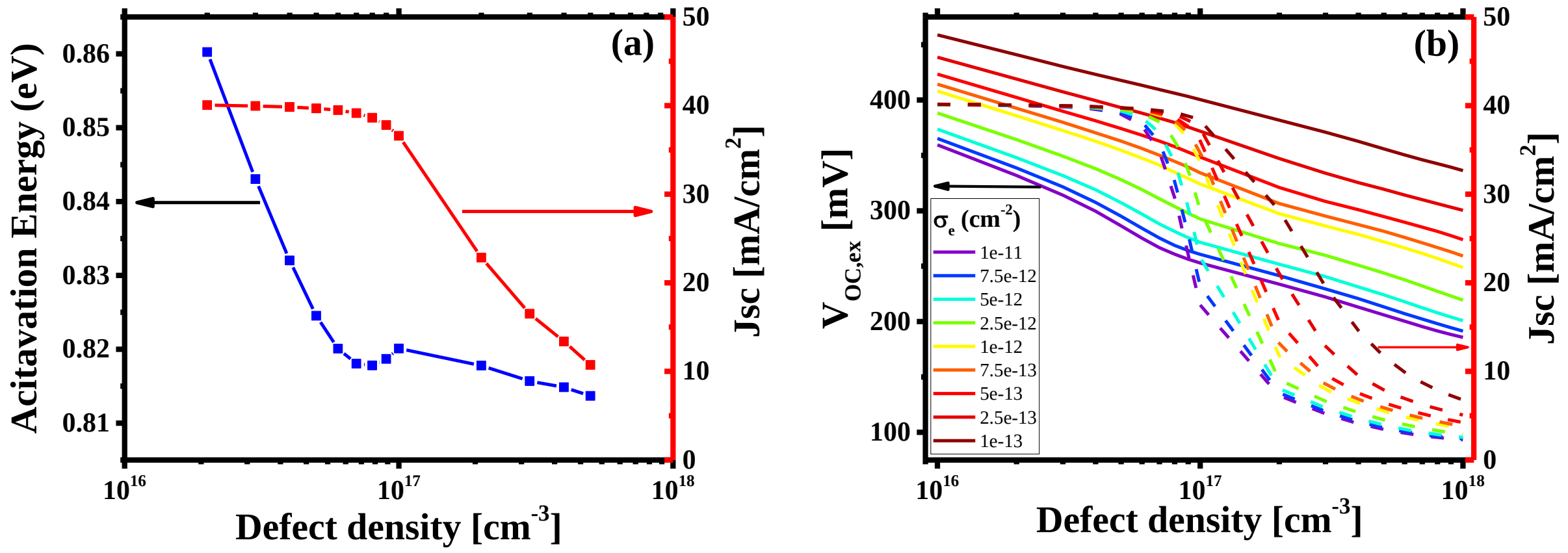

***Figure S8***. (a) Activation energy ($E_a$) and short-circuit current density ($J_{sc}$) as a function of defect density in the defective layer. The electron capture cross-section values were adjusted in each case to obtain similar $V_{OC,ex}$ at 300K in order to obtain the $E_a$ values. (b) $V_{OC,ex}$ and $J_{sc}$ as a function of defect density in the defective layer.

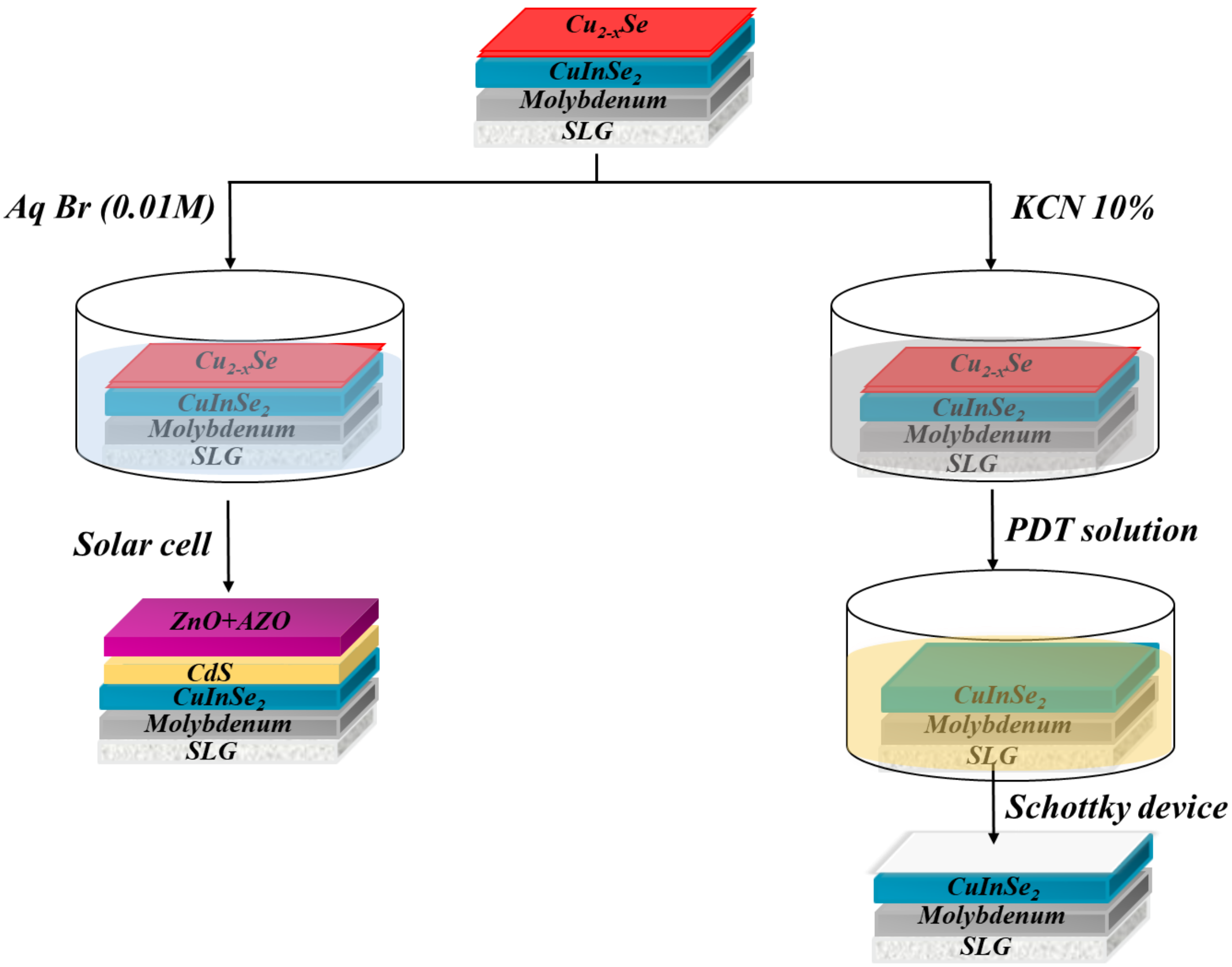

***Figure S9.*** *A schematic diagram showing the procedure used for bromine etching and post deposition treatment (Zn, Cd and S-PDT). The secondary phase $Cu_{2-X}Se$ are etched from Cu-rich absorber using 10% KCN for 5 minutes, followed by PDT of absorbers in either ammoniac Zn or Cd or S solution at 80° C . Finally, aluminum is deposited on the absorber to make schottky contact. In case of bromine etching buffer (CdS) and window (aluminum doped zinc oxide i.e. AZO) followed by Nickle Aluminum grids are deposited.*